%% file: main.tex
\documentclass[10pt,twocolumn,twoside]{IEEEtran}

\usepackage{amsmath}
\usepackage{amsfonts}
\usepackage{amsthm}
\usepackage{graphicx}
\usepackage{caption}
\usepackage{subcaption}
\usepackage{mathtools}
\usepackage{array}
\usepackage{verbatim}
\usepackage{mathtools}
\usepackage{blkarray, bigstrut}
\usepackage{color}
\usepackage{fancyhdr}

\newtheorem{theorem}{Theorem}
\newtheorem{corollary}{Corollary}
\newtheorem{defn}{Definition}
\newtheorem{lemma}{Lemma}
\newtheorem{ex}{Example}

\DeclarePairedDelimiter{\ceil}{\lceil}{\rceil}
\DeclarePairedDelimiter\floor{\lfloor}{\rfloor}
\DeclareMathOperator*{\argmax}{arg\,max}
\DeclareMathOperator*{\argmin}{arg\,min}

\renewcommand{\headrulewidth}{0pt}
\title{\LARGE \bf The Impact of Information in Distributed Submodular Maximization}
\author{David Grimsman, Mohd. Shabbir Ali, Jo\~{a}o P. Hespanha and Jason R. Marden \thanks{D. Grimsman (\texttt{davidgrimsman@umail.ucsb.edu}, Corresponding Author), J. R. Marden (\texttt{jmarden@ece.ucsb.edu}), and J. P. Hespanha (\texttt{hespanha@ece.ucsb.edu}) are with the Department of Electrical and Computer Engineering, UC Santa Barbara and can be reached at Room 4155, Harold Frank Hall, ECE Dept., University of California, Santa Barbara, CA 93106 or by phone at +1 805-893-5364.}
	\thanks{M. S. Ali (\texttt{mdshabbirali88@gmail.com}) is with Orange Labs in Paris, France, and can be reached at +33 751078397. He was sponsored by NetLearn ANR project (ANR-13-INFR-004).}
	\thanks{This work is supported by ONR Grants \#N00014-17-1-2060 and \#N00014-16-1-2710 and NSF Grants \#ECCS-1638214 and \#ECCS-1608880.}
	\thanks{A preliminary version of this work appeared in \cite{grimsmanimpact}}}

\begin{document}

\maketitle
\thispagestyle{fancy}
\renewcommand{\headrulewidth}{0pt}
\fancyhead{}
\fancyhead[R]{\thepage}
\fancyfoot{}
\fancyfoot[C]{ \footnotesize Copyright (c) 2019 IEEE. This is the author's version of an article that has been published in Transactions on Control of Network Systems. Changes were made to this version by the publisher prior to publication. The final version of record is available at http://dx.doi.org/10.1109/TCNS.2018.288900}
\input{src/abstract}
\input{src/intro}
\input{src/model}
\input{src/efficiency}
\input{src/structures}
\input{src/conclusion}
\bibliographystyle{plain}
\bibliography{util/refs}
\input{src/appendix}
\input{src/authors}

\end{document}

%% file: src/abstract.tex
\begin{abstract}
	    The maximization of submodular functions is an NP-Hard problem for certain subclasses of functions, for which a simple greedy algorithm has been shown to guarantee a solution whose quality is within 1/2 of the optimal. When this algorithm is implemented in a distributed way, agents sequentially make decisions based on the decisions of all previous agents. This work explores how limited access to the decisions of previous agents affects the quality of the solution of the greedy algorithm. Specifically, we provide tight upper and lower bounds on how well the algorithm performs, as a function of the information available to each agent. Intuitively, the results show that performance roughly degrades proportionally to the size of the largest group of agents which make decisions independently. Additionally, we consider the case where a system designer is given a set of agents and a global limit on the amount of information that can be accessed. Our results show that the best designs partition the agents into equally-sized sets and allow agents to access the decisions of all previous agents within the same set.
\end{abstract}

%% file: src/intro.tex
\section{Introduction}

The optimization of submodular functions is a well-studied topic due to its application in many common engineering problems. Examples include 
information gathering~\cite{krause2007near},
maximizing influence in social networks~\cite{kempe2003maximizing},
image segmentation in image processing~\cite{kohli2009p3},
multiple object detection in computer vision~\cite{barinova2012detection},
document summarization~\cite{lin2011class},
path planning of multiple robots~\cite{singh2007efficient},
sensor placement~\cite{krause2009simultaneous,marden2016}, and
resource allocation in multi-agent systems~\cite{marden2016}. The key thread in these problems is that each exhibits some form of a ``diminishing returns" property, e.g., adding more sensors to a sensor placement problem improves performance, but every additional sensor marginally contributes less to the overall performance as the number of sensors increases. Any problem exhibiting such behavior can likely be formulated as a submodular optimization problem.

While polynomial algorithms exist to solve submodular minimization problems, \cite{grotschel1981ellipsoid,iwata2001combinatorial,schrijver2000combinatorial}, maximization has been shown to be NP-Hard for important subclasses of submodular functions~\cite{lovasz1983submodular}. Thus a tremendous effort has been placed on developing fast algorithms that approximate the solution to the submodular maximization problem~\cite{nemhauser1978analysis,fisher1978analysis,minoux1978accelerated,buchbinder2015tight,vondrak2008optimal,sviridenko2004note,qu2016distrib}. A resounding message from this extensive research is that very simple algorithms can provide strong guarantees on the quality of the approximation. 

The seminal work in \cite{fisher1978analysis} demonstrates that a greedy algorithm provides a solution that is within $1/2$ of the quality of the optimal solution. In fact, more sophisticated algorithms can often be derived for certain classes of submodular maximization problems that push these guarantees from $1/2$ to $1-1/e$ \cite{nemhauser1978analysis,calinescu2007maximizing,filmus2012power}. Progress beyond this level of suboptimality is not possible in general, because it was also shown that no polynomial-time algorithm can achieve a higher guarantee than $(1 - 1/e)$, unless $P=NP$ \cite{feige1998threshold}.

One appealing trait of the greedy algorithm is that it can be implemented in a distributed way while still maintaining the $1/2$ performance guarantee. In the distributed greedy algorithm, each agent in the set sequentially selects its choice by greedily optimizing the global objective function conditioned on the decisions of the previous agents. However, this requires agents to have full access to the decisions of all previous agents. In addition, each agent must have access to the value of the global objective function for previous agents' decisions, plus any choice in its own decision set. In many cases, these informational demands may be costly or infeasible.

Research has therefore begun to explore how limited information can impact the performance of the greedy algorithm on distributed submodular maximization problems. For example, \cite{marden2016} focuses on the submodular resource allocation problem, modeled as a game played among agents. The resulting Nash equilibria have the familiar $1/2$ performance guarantee, however it is shown that when information is limited to be local instead of global, the performance guarantee degrades to $1/n$, where $n$ is the number of agents. The work in \cite{mirzasoleiman2013distributed} formulates the problem of selecting representative data points from a large corpus as a submodular maximization problem. In order to perform the optimization in a distributed way, agents are partitioned into sets, where the full greedy algorithm is performed among agents within a set, while no information is transferred between sets. In this setting, the paper shows that the algorithm performance is worse than $1/2$, even when a preprocessing algorithm is used to intelligently assign decision sets to each agent. Other work in \cite{qu2016distrib} discusses the role of information in the task assignment problem. It is shown that the distributed greedy algorithm can be implemented asynchronously, with convergence in a finite number of steps. Additionally, when agent action sets are based on spatial proximity, agents need only consider local information to achieve the $1/2$ bound. Finally, the work in \cite{gharesifard2016distributed} studies the performance of the distributed greedy algorithm when an agent can only observe a local subset of its predecessors. It is shown that localizing information, particularly when agents are partitioned from each other, leads to a degradation in performance. For instance, in the case where agents are partitioned into sets, performing the full greedy algorithm within the set and obtaining no information outside the set, the performance degrades proportionally to the number of sets in the partition.

This paper more closely relates to the work done in~\cite{gharesifard2016distributed} in evaluating informational constraints. We leverage a similar model and seek to find how limiting which decisions an agent can access impacts the overall performance of the distributed greedy algorithm. We also consider the scenario where a system designer is given a set of agents and a global limit on the amount of information that can be accessed, and seek to find the best policies to ensure the highest performance possible. 

More specifically, the contributions of this paper are the following results:
\begin{enumerate}
	\item Theorem~\ref{thm:eff} gives lower and upper bounds on worst-case performance of the greedy algorithm on any submodular function for any given set of constraints. The bounds show that worst-case performance (roughly) degrades proportionally to the size of the largest group of agents which make decisions independently.
	\item Theorem~\ref{thm:struct} shows the best performance of the greedy algorithm that a system designer can achieve with a fixed number of agents and constraints. The results show that when information is costly, the best system design is to partition the agents into equal sets, and have agents in each set execute the full greedy algorithm.
\end{enumerate}
The remainder of this paper is dedicated to proving and discussing these two theorems.

%% file: src/model.tex
\section{Model}
\label{sec:model}

This paper focuses on a distributed algorithm for solving submodular maximization. To that end, let $S$ be a set of elements and $f: 2^S \to \mathbb{R}_{\geq 0}$ have the following properties:
\begin{itemize}
	\item \emph{Normalized}: $f(\emptyset) = 0$.
	\item \emph{Monotonic}: For $A \subseteq B \subseteq S$, $f(A) \leq f(B)$.
    \item \emph{Submodular}: For $A \subseteq B \subset S$ and $x \in S \setminus B$, the following holds:
    \begin{equation}
    \label{eq:submod}
        f(A \cup \{x\}) - f(A) \geq f(B \cup \{x\}) - f(B).
    \end{equation}
\end{itemize}
For simplicity, we will refer to a function with all three above properties merely as submodular.

In this paper we focus on distributed approaches to submodular optimization where there are a set of decision-making agents $N = \{1, 2, \dots, n\}$, and each agent $i$ is associated with an \emph{action set} $X_i \subseteq 2^S$. Notationally, we define $X = X_1 \times \cdots \times X_n$ as the family of action sets, an \emph{action} for agent $i$ as $x_i \in X_i$, and an \emph{action profile} as $x \in X$. We also overload the notation of $f$ to allow multiple actions as inputs: $f(x_i, x_j) := f(x_i \cup x_j)$, $f(x) := f(\bigcup_i x_i)$ and $f(x_{a:b}) := f \left( \bigcup_{a \leq i \leq b} x_i \right)$ for $a \leq b$. The submodular maximization problem addressed in this work is to find
\begin{equation}
\label{eq:max}
    x^{\rm opt} \in \max_{x \in X} f(x).
\end{equation}
As shown in~\cite{gharesifard2016distributed}, this type of constraint where we choose from a family of subsets $X_i$ is also referred to in the literature as a partition matroid constraint.

We next present two relevant problems that can be modeled accordingly. This serves to give a scope and relevance to the model, as well as provide an example that will be leveraged throughout the rest of the paper.

\begin{ex}[Vehicles target assignment problem \cite{arslan2007autonomous}]\label{ex:vta}
	Consider the classic vehicles target assignment problem where there are a collection of targets ${\cal T}$ and each target $t \in {\cal T}$ has an associated value $v_t \geq 0$.  Further, there exists a collection of $n$ agents, and each agent $i$ is associated with a success probability $p_i \in [0,1]$ and a set of possible assignments $X_i \subseteq 2^{\cal T}$.  The agents make decisions to reach a feasible allocation of agents to targets $x = (x_1, \dots, x_n) \in X_1 \times \dots \times X_n$ that optimizes a system-level performance metric of the form:
	\begin{equation}\label{eq:wta}
	W(x) = \sum\nolimits_{t \in \cup_i x_i} v_t \left( 1- \prod\nolimits_{i:t \in x_i}(1-p_i) \right). 
	\end{equation}
	Note that the objective function given in \eqref{eq:wta} is submodular, as $W$ can be expressed as a function of the form $W: 2^S \to \mathbb{R}_{\geq 0}$ for an appropriate choice of the domain set $S$, i.e., $S = N \times 2^{\cal T}$ and the action sets can be expressed as disjoint sets in $S$.   
\end{ex}

\begin{figure}
	\begin{subfigure}[b]{0.48\textwidth}
		\centering
		\includegraphics[scale=0.5]{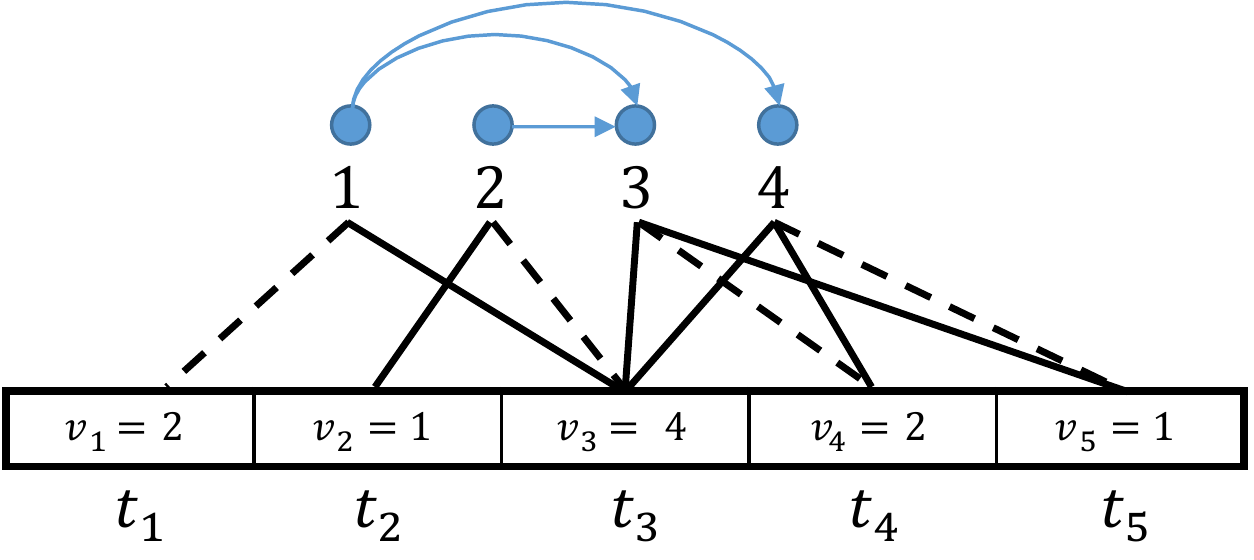}
		\caption{The setup of a weighted set cover problem. The targets are $\mathcal{T} = \{t_1, ..., t_5\}$, each represented by a box, and each with a corresponding value. The available choices to each agent are represented by the black lines (both dotted and solid) - for instance $X_1 = \{\{t_1\}, \{t_3\}\}$, $X_2 = \{\{t_2\}, \{t_3\}\}$, etc. The dashed lines represent an optimal set of choices. The goal for the agents is to maximize $W(x)$ in \eqref{eq:wta22}. Using the generalized distributed algorithm (i.e., agents choose according to \eqref{eq:choice2}), agent 1 chooses $t_3$, since $v_3 > v_1$. Then, agent 2, who (according to the graph) does not know that agent 1 has chosen $t_3$, also chooses $t_3$, since $v_3 > v_2$. Agent 3 observes that agents 1 and 2 have both chosen $t_3$, so it chooses $t_4$, since $v_4 > v_5$. Finally, agent 4, observing that agent 1 has chosen $t_3$ (but not that agent 3 has chosen $t_4$), chooses $t_4$, since $v_4 > v_5$. These results are summarized in the table below.\\}
		\label{fig:wscgraph}
	\end{subfigure}
	\begin{subfigure}[b]{0.48\textwidth}
		\centering
		\resizebox{0.9\textwidth}{!}{%
			\begin{tabular}{m{0.36\textwidth}|m{0.06\textwidth}|m{0.06\textwidth}|m{0.06\textwidth}|m{0.06\textwidth}|m{0.1\textwidth}}
				Algorithm & $x_1^{\rm sol}$ & $x_2^{\rm sol}$ & $x_3^{\rm sol}$ & $x_4^{\rm sol}$ & $f(x^{\rm sol})$\\ \hline \hline
				Optimal & $\{t_1\}$ & $\{t_3\}$ & $\{t_4\}$ & $\{t_5\}$ & 9 \\ \hline
				Distributed Greedy & $\{t_3\}$ & $\{t_2\}$ & $\{t_4\}$ & $\{t_5\}$ & 8 \\ \hline
				Generalized Distributed Greedy & $\{t_3\}$ & $\{t_3\}$ & $\{t_4\}$ & $\{t_4\}$ & 6 \\ \hline
		\end{tabular}}
		\caption{For the weighted set cover problem outlined above, this table shows the agents' decisions in an optimal case, the case where the distributed greedy algorithm is used (agents choose according to \eqref{eq:choice1}) and the case where the generalized distributed algorithm is used (agents choose according to \eqref{eq:choice2}, constrained to the graph shown above). The difference between the distributed greedy algorithm and the generalized version can be seen in the choices of agents 2 and 4. Agent 2 chooses $t_3$ when it can observe that $t_4$ has already been chosen by agent 1, otherwise it chooses $t_4$. Likewise, agent 4 chooses $t_5$ only when it knows that $t_4$ has already been selected. Therefore, as the informational constraints grow, the solution quality decreases. As a note, in this case we see that $\gamma(f, X, G) = 6/9$.}
	\end{subfigure}
	\caption{An instance of the weighted set cover problem and the performance of the greedy algorithm in solving it.}
	\label{fig:wsc}
\end{figure}

\begin{ex}[Weighted set cover problem \cite{gairing2009covering}]
	Consider the subset of the vehicles target assignment problem where $p_i = 1$ for all $i$. Then, the objective function takes on the form
	\begin{equation}\label{eq:wta22}
	W(x) = \sum\nolimits_{t \in \cup_i x_i} v_t. 
	\end{equation}
	Note that \eqref{eq:wta22} can now be expressed by a submodular function with domain $S = {\cal T}$.  An instance of such a problem is shown in Figure \ref{fig:wsc}.
\end{ex}

We will henceforth focus on submodular functions of the form $f:2^S \to \mathbb{R}_{\geq 0}$, where $X_i \subseteq 2^S$, without explicitly highlighting the structure of $S$ and the action sets $X_1, \dots, X_n$. Furthermore, we will rely on the weighted set cover problem in several of the forthcoming proofs.

\section{The Greedy Algorithm}

One of the most well-studied algorithms to solve the submodular maximization problem is the greedy algorithm. This algorithm requires agents to make decisions sequentially
\footnote{Although we state that the agents must choose sequentially, the real restriction is that the flow of information in the system is acyclic. Accordingly, if the agents select their decisions with regards to another process, e.g., a synchronous best reply process, they will still arrive at the same solution. Once agent 1 has chosen, it will not make a different choice regardless of order, and agent 2 will not switch after agent 1 has decided, etc. See \cite{gharesifard2016distributed} for more details).}
, so without loss of generality we impose an ordering on the agents according to their labels, i.e., agent 1 chooses first, agent 2 chooses second, etc. Agent $i$ makes its choice $x^{\rm sol}_i \in X_i$ based on the following rule:
\begin{equation}
\label{eq:choice1}
    x^{\rm sol}_i \in \argmax_{x_i \in X_i} f\left(x_i, x_{1:i-1}^{\rm sol}\right),
\end{equation}
In words, each agent $i$ selects the action that would optimize the objective function $f$ given knowledge of the action choices of the previous agents and the global objective function $f$.

It is well-known that for any submodular $f$, any set of action spaces $X$, and any order of agents, the quality of the resulting solution $x^{\rm sol}$ derived from the distributed greedy algorithm compared to the optimal solution $x^{\rm opt}$ satisfies
\begin{equation}
   \frac{f(x^{\rm sol})}{f(x^{\rm opt})} \geq \frac{1}{2},
\end{equation}
where $x^{\rm sol} = (x_1^{\rm sol} , \cdots , x_n^{\rm sol})$. In other words, the quality of the solution is within $1/2$ that of the optimal~\cite{nemhauser1978analysis}. For special classes of submodular functions, and additional constraints placed on $X_1, \dots, X_n$ (for instance if $X_1 = \cdots = X_n$), \cite{nemhauser1978analysis} also shows that the solution to the greeedy lies within $1 - 1/e \approx 63\%$ of the optimal. 

In~\eqref{eq:choice1}, agent $i$ must have access to the decisions of all previous agents. However, there are many applications where this level of informational demand may be impractical. A more generalized version of the distributed greedy algorithm is proposed in \cite{gharesifard2016distributed}, where each agent $i$ makes its choice using the following rule:
\begin{equation}
\label{eq:choice2}
    x^{\rm sol}_i \in \argmax_{x_i \in X_i} f\left(x_i, x_{\mathcal{N}_i}^{\rm sol}\right),
\end{equation}
where $\mathcal{N}_i \subseteq \{1, ..., i-1\}$, $\mathcal{N}_1 = \emptyset$ and $x_{{\cal N}_i} = \cup_{j \in \mathcal{N}_i} x_j$. The sets $\mathcal{N}_1, ..., \mathcal{N}_n$ characterize the informational constraints of the agent in the sense that $\mathcal{N}_i$ is the set of agents whose choices agent $i$ can access when making its own action decision. The central topic of discussion here is how the structure of $\mathcal{N}_1, ..., \mathcal{N}_n$ impacts the performance guarantees associated with this generalized greedy algorithm. Note that \eqref{eq:choice2} is simply one decision rule that could be used by the agents. Analysis of whether this rule is optimal among all possible decision rules given the local information is the topic of ongoing research.

It is helpful to model the informational constraints as a graph $G = (V, E)$, where $V$ is a set of nodes and $E \subseteq V \times V$ is a set of directed edges between nodes. In this scenario each node is an agent (and thus we use the terms interchangeably) and each edge $(j, i)$ implies that $j \in \mathcal{N}_i$, i.e., $\mathcal{N}_i$ is the set of in-neighbors for vertex $i$. Since there is an imposed ordering on the vertices, and the agents choose sequentially, the set $\mathcal{G} := \{G = (V, E) : (i, j) \in E \implies i < j\}$ is the set of admissible graphs that correspond to a set of informational constraints.

The solution of the algorithm defined by \eqref{eq:choice2} is denoted $x^{\rm sol}(f, X, G)$ and the optimal decisions as $x^{\rm opt}(f, X)$ to explicity highlight the dependence of \eqref{eq:choice2} on the graph $G$. We define the efficiency guarantees associated with this algorithm as:
\begin{equation}
	\gamma(f, X, G) := \frac{f(x^{\rm sol}(f, X, G))}{f(x^{\rm opt}(f, X))}.
\end{equation}
Note that $x^{\rm sol}(f, X, G)$ could in fact be a set when \eqref{eq:choice2} is not unique, so we write $f(x^{\rm sol}(f, X, G))$ with the understanding that $f$ is evaluated at the worst possible candidate solution, i.e., $\min_{x \in x^{\rm sol}(f, X, G)} f(x)$.

One goal of this paper is to characterize the efficiency guarantees associated with this more generalized version of the distributed greedy algorithm for any submodular function and action sets $X$. To that end, we define
\begin{equation}
    \gamma(G) := \inf_{f, X} \gamma(f, X, G),
\end{equation}
In words, $\gamma(G)$ is the worst-case efficiency for any $f$ and family of sets $X_1,...,X_n$ as defined above, given the informational constraints among the agents represented by $G$.

%% file: src/efficiency.tex
\section{Efficiency Bounds}
\label{sec:eff}

In this section we present lower and upper bounds for the worst-case efficiency $\gamma(G)$ based on the structure of the graph $G \in \mathcal{G}$. We begin with some preliminaries from graph theory, and then present the bounds.

\subsection{Preliminaries}\label{subsec:graph}

\begin{figure*}
	\begin{subfigure}[t]{0.61\textwidth}
		\centering
		\includegraphics[scale=0.5]{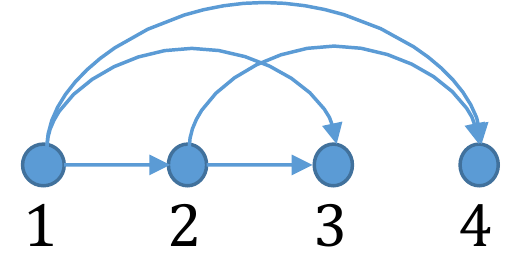}
		\caption{In this graph, there are 4 cliques of size 1 (one for each node), 5 cliques of size 2 (one representing each edge), and 2 cliques of size 3 (the sets $\{1,2,3\}$ and $\{1, 2, 4\}$). Thus $\omega(G) = 3$. A minimum clique cover is $\{1, 3\}, \{2, 4\}$, so $k(G)=2$. The maximum independent set is $\{3, 4\}$, thus $\alpha(G)=2$. Since $\alpha(G) = k(G) = 2$, we know that $\alpha^*(G) = k^*(G) = 2$. Appendix \ref{app:exupper} shows that $\gamma(G) = 1/2$, making it a graph that meets the upper bound for Theorem \ref{thm:eff} (see Section \ref{subsec:ex}). Lastly, it is also an example of a graph without the Sibling Property (see Section \ref{subsec:sp}), since no such $w$ exists from Definition \ref{def:sp}.\\}
		\label{fig:graph}
	\end{subfigure}\hfill
	\begin{subfigure}[t]{0.37\textwidth}
		\centering
		\includegraphics[scale=0.5]{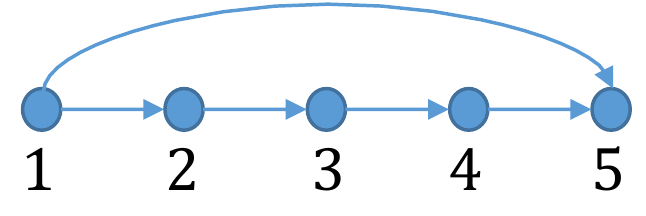}
		\caption{A graph where $\alpha(G) = 2$, $k(G) = 3$, $\alpha^*(G) = k^*(G) = 2.5$, and $z = [1/2, 1/2, 1/2, 1/2, 1/2]^T$ maximizes \eqref{eq:fracindlp}. As a note, this is the graph with the fewest number of nodes and edges such that $\alpha(G) \neq k(G)$. This is also a graph with the Sibling Property (see Section \ref{subsec:sp}), since for maximum independent set $\{2, 4\}$, $2 \in \mathcal{N}_3$, so $w=3$ from Definition \ref{def:sp}.}
		\label{fig:frac_ind}  
	\end{subfigure}
	\caption{Two example graphs showcasing the graph properties defined in Section \ref{subsec:graph}. These graphs will be referred to throughout the paper to illustrate the tightness of bounds in Theorem \ref{thm:eff} and to illustrate the Sibling Property (see Section \ref{subsec:sp}).}
	\label{fig:ex}
\end{figure*}

For all definitions in this section, we assume that $G = (V, E)$ is a general directed graph. We begin with cliques: a \emph{clique} is a set of nodes $C \subseteq V$ such that for every $i, j \in C$, either $(i, j) \in E$ or $(j, i) \in E$. The \emph{clique number} $\omega(G)$ is the number of nodes in the largest clique in $G$. We denote by $K(G)$ the set of all cliques in $G$. A \emph{clique cover} is a partition on $V$ such that the nodes in each set of the partition form a clique. The \emph{clique cover number} $k(G)$ is the minimum number of sets needed to form a clique cover of $G$. For an example, see Figure \ref{fig:graph}.

Another important notion in graph theory is that of independence. An \emph{independent set} $J \subseteq V$ is a set of vertices such that $v_1, v_2 \in J$ implies $(v_1, v_2), (v_2, v_1) \notin E$. A \emph{maximum independent set} is an independent set of $G$ such that no other independent set has more vertices. The \emph{independence number} $\alpha(G)$ is the number of nodes in the largest independent set in $G$. For an example, see Figure \ref{fig:graph}. 

The work in \cite{godsil2013algebraic} equivalently characterizes the independence number as the solution to an integer linear program~\footnote{It is actually the chromatic number and clique number that are defined this way in \cite{godsil2013algebraic}. However, using graph complementarity, it is an easy extension to show that the solution to the linear program in \eqref{eq:indlp} yields a maximum independent set.}. Let $W \in \mathbb{R}^{|K(G)| \times n}$ be the binary matrix whose rows are indicator vectors for the cliques in $G$. In other words, $W_{ij} = 1$ if node $j$ belongs to clique $i$ in $G$, and 0 otherwise. Note that $W$ also includes cliques of size 1 (the individual nodes). Then $\alpha(G)$ is given by
\begin{equation}
\label{eq:indlp}
	\begin{aligned}
		& \max_z
		& & z^T \textbf{1} \\
		& \text{subject to}
		& & Wz \leq \textbf{1} \\
		& & & z \in \mathbb{Z}^n \geq \textbf{0}.
	\end{aligned}
\end{equation}
It is similarly shown that $k(G)$ is characterized by the dual to this problem, implying that $\alpha(G) \leq k(G)$. As an example, for the graph in Figure \ref{fig:graph},
\begin{equation}
	W=
	\begin{blockarray}{*{4}{c} l}
		\begin{block}{*{4}{>{$\footnotesize}c<{$}} l}
			Node 1 & Node 2 & Node 3 & Node 4 \\
			\end{block}
			\begin{block}{[*{4}{c}]>{$\footnotesize}l<{$}}
			1 & 0 & 0 & 0 \bigstrut[t] & $\{1\}$ \\
			0 & 1 & 0 & 0 & $\{2\}$ \\
			0 & 0 & 1 & 0 & $\{3\}$ \\
			0 & 0 & 0 & 1 & $\{4\}$ \\
			1 & 1 & 0 & 0 & $\{1, 2\}$ \\
			1 & 0 & 1 & 0 & $\{1, 3\}$ \\
			1 & 0 & 0 & 1 & $\{1, 4\}$ \\
			0 & 1 & 1 & 0 & $\{2, 3\}$ \\
			0 & 1 & 0 & 1 & $\{2, 4\}$ \\
			1 & 1 & 1 & 0 & $\{1, 2, 3\}$ \\
			1 & 1 & 0 & 1 & $\{1, 2, 4\}$ \\
		\end{block}
	\end{blockarray}
\end{equation}
Using this $W$ in \eqref{eq:indlp}, it is straightforward to show that the optimal solution is $z = [0, 0, 1, 1]^T$, i.e., $\alpha(G)=2$, and the maximum independent set is $\{3, 4\}$.

Note by defintion that $\alpha(G)$ and $k(G)$ are always positive integers. However, in many applications, it is helpful to consider a real-valued relaxation on these notions: this is the motivation for fractional graph theory \cite{godsil2013algebraic}. Here we leverage the \emph{fractional independence number} $\alpha^*(G)$, which we define as the real-valued relaxation to \eqref{eq:indlp}: \footnote{Another defintion of fractional independence exists in the literature (see \cite{arumugam2007fractional}), which was created to preserve certain properties of graph independence (such as nested maximality), but has not been shown to preserve $\alpha^*(G) = \omega^*(\bar{G})$, where $\bar{G}$ is the complement graph of $G$ and $\omega^*(G)$ is the fractional clique number of $G$.}

\begin{equation}
\label{eq:fracindlp}
	\begin{aligned}
	\alpha^*(G) := & \max_z
	& & z^T \textbf{1} \\
	& \text{subject to}
	& & Wz \leq \textbf{1} \\
	& & & z \geq \textbf{0}.
	\end{aligned}
\end{equation}  
Likewise, $k^*(G)$, the \emph{fractional clique cover number} of $G$, can be defined by its dual

\begin{equation}
\label{eq:fraccliquecovlp}
	\begin{aligned}
	k^*(G) := & \min_y
	& & y^T \textbf{1} \\
	& \text{subject to}
	& & W^Ty \geq \textbf{1} \\
	& & & y \geq \textbf{0}.
\end{aligned}
\end{equation}
In accordance with the Strong Duality of Linear Programming, it follows that:
\begin{equation}
\label{eq:fracprop}
	\alpha(G) \leq \alpha^*(G) = k^*(G) \leq k(G).
\end{equation}
An example of a graph where the independence number differs from the fractional independence number is found in Figure~\ref{fig:frac_ind}.

\subsection{Result}

We now present results regarding the quality of the solution provided by the generalized distributed greedy algorithm subject to the informational constraints represented by a graph $G$. We show that the performance degrades proportionally to the fractional independence number of $G$ \footnote{In comparison to the bounds shown in \cite{gharesifard2016distributed}, the bounds shown in our work are tighter in all cases. In fact, except in certain corner cases (for example, both bounds are the same on a full clique), our results are strictly tighter.}. 

\begin{theorem}
	\label{thm:eff}
	For any graph $G \in \mathcal{G}$, 
	\begin{equation}
		\frac{1}{\alpha^*(G)} \geq \gamma(G) \geq \frac{1}{\alpha^*(G) + 1}.
	\end{equation}
	The upper bound shows that it is impossible to construct a graph $G$ such that the greedy algorithm's performance is better than $1/\alpha^*(G)$ for all possible $f$ and $X$. Likewise, the lower bound means that no $f$ and $X$ can result in a performance lower than $1/(\alpha^*(G) + 1)$.  
\end{theorem}

The formal proof for this theorem is given in Section \ref{subsec:thmeff}, but here we give a brief outline of the argument. For the upper bound, we develop a canonical $f$ and $X$, which are dependent on the cliques in $G$. Then we show that for this example, $\gamma(f, X, G)$ is the inverse of the solution to \eqref{eq:fracindlp}, i.e., $\alpha^*(G)$. The lower bound is found by leveraging the properties of submodularity and monotonicity, showing that the highest lower bound requires solving \eqref{eq:fraccliquecovlp}. Then, using \eqref{eq:fracprop}, we show that Theorem \ref{thm:eff} holds.

\subsection{Examples}
\label{subsec:ex}

Theorem \ref{thm:eff} shows lower and upper bounds on $\gamma(G)$, but we have not shown whether either of these bounds is tight. There exist graphs for which $f$ and $X$ can be chosen to meet the lower bound, and there also exist graphs whose lower bound can be proven to meet the upper bound. In this section, we provide an example of each.

\begin{figure}
	\centering
	\includegraphics[scale=0.5]{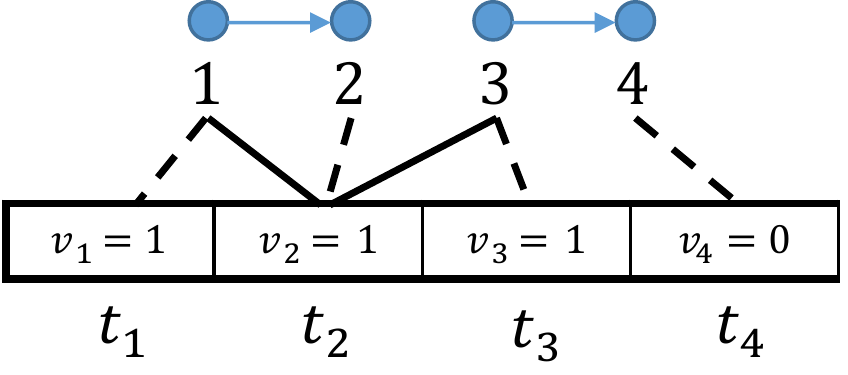}
	\caption{An example of a graph $G$ where $\gamma(G) = 1/(\alpha^*(G) + 1)$, and an instance of a weighted set cover problem using the same notation as in Figure \ref{fig:wsc}. Here $\alpha(G) = \alpha^*(G) = 2$, and we can see that $f(x^{\rm opt}) = 3$. The worst-case results from the generalized distributed greedy algorithm occur when $x_1^{\rm sol}=x_2^{\rm sol}=x_3^{\rm sol}=\{t_2\}$, and therefore $f(x^{\rm sol}) = 1$. This means $\gamma(f, X, G) = 1/(\alpha^*(G) + 1) = 1/3$, so the lower bound in Theorem \ref{thm:eff} is tight for this graph.}
	\label{fig:tightlower}
\end{figure}

\begin{ex}
	The weighted set coverage problem presented in Figure \ref{fig:tightlower} is an example showing that the lower bound from Theorem \ref{thm:eff} is tight. For this graph $G$, $\alpha(G) = \alpha^*(G) = 2$. As shown, $\gamma(f, X, G) = 1/(\alpha(G) + 1)$. Since $\gamma(G) \leq \gamma(f, X, G)$ by definition, it follows that $\gamma(G) = 1/(\alpha^*(G) + 1)$.
\end{ex}

\begin{ex}
\label{ex:upper}
	The graph $G$ in Figure \ref{fig:graph} is an example where the upper bound from Theorem \ref{thm:eff} is tight. Here $\alpha^*(G) = 2$, and it is shown in Appendix \ref{app:exupper} for this graph that no $f$ and $X$ can be constructed to give a worse efficiency than $1/2$.
\end{ex}

\subsection{Proof for Theorem \ref{thm:eff}}
\label{subsec:thmeff}

Before starting the proof, we introduce some notation. Let $A \subseteq N$ and $x_A = \bigcup_{j \in A} x_j$. Then we define
\begin{equation}
\label{eq:marg}
\Delta(x_i|x_A) := f(x_i, x_A) - f(x_A)
\end{equation}
to be the marginal contribution of the choice of agent $i$ given the choices of the agents in set $A$. One property to note about $\Delta$ is the following holds when $G$ is a clique:
\begin{equation} \label{eq:margclique}
	\sum\nolimits_i \Delta\left(x_i, x_{1:i-1}\right) = f(x).
\end{equation}

We now begin the theorem proof. We first show the upper bound by constructing a canonical example $f$ and $X$, which can be applied to any $G$, and show that $\gamma(f, X, G) = 1/\alpha^*(G)$. Then we prove the lower bound leveraging the properties of submodularity and monotonicity.

Consider a base set of elements $S = \{u_1, ..., u_n, v_1, ..., v_n\}$ and let $X_i = \{\{u_i\}, \{v_i\}\}$ for all $i$. We endeavor to design submodular $f$ so that for all $i$,$x_i^{\rm opt} = \{v_i\}$ and in the worst case $x^{\rm sol}_i = \{u_i\}$. Let $f: 2^S \to \mathbb{R}_{\geq0}$ have the following restrictions:
\begin{enumerate}
	\item $f\left(\{u_i\}\right) = f\left(\{v_i\}\right) \geq 0$.
	\item $f(\{u_1, ..., u_n\}) \leq 1$.
	\item $\Delta (\{u_i\} | \cup_{j \in A} \{u_j\}) = f(\{u_i\})$ for any $A \subseteq \mathcal{N}_i$.
	\item $\Delta (\{v_i\} | B) = f(\{v_i\})$ for all $i \in N$ and $B \subseteq S \setminus \{v_i\}$.
\end{enumerate}

First note that Restrictions 2 and 3 hold if and only if for all $c \in K(G)$
\begin{equation}
\label{eq:effubconst}
f \left(\cup_{i \in c} \{u_i\}\right) = \sum\nolimits_{i \in c} f(\{u_i\}) \leq 1.
\end{equation}

If one sets values for all $f(\{u_i\})$ which meet this requirement, a full function on $2^S$ can be created with a simple extension. Let $A \subseteq S$, then define
\begin{equation}
\label{eq:ubsubmodfun}
	f(A) = \max\left\{1, \sum\nolimits_{u_i \in A} f(\{u_i\})\right\} + \sum\nolimits_{v_i \in A}f(\{v_i\}).
\end{equation}

We claim the function $f$ created in this way is normalized, monotone, and submodular. For $A \subseteq B \subset S$, we know by Restriction 4 that for $v_i \notin B$, $f(A,\{v_i\}) - f(A) = f(B, \{v_i\}) - f(B) = f(\{v_i\})$. For $u_i \notin B$, it should be clear from \eqref{eq:ubsubmodfun} that $f(A, \{u_i\}) - f(A) \geq f(B, \{u_i\}) - f(B)$. We also see that $f(B) \geq f(A)$ (monotone) and $f(\emptyset) = 0$ (normalized). 

Restrictions 1, 3, and 4 imply that, since agents are choosing according to \eqref{eq:choice2}, every agent is choosing between equally desirable choices. \color{black} However, it should be clear from Restriction 4 that $x_i^{\rm opt} = \{v_i\}$, and in the worst case $x^{\rm sol}_i = \{u_i\}$. Therefore,
\begin{align}
\gamma(f, X, G) = \frac{f(\{u_1\}, ..., \{u_n\})}{f(\{v_1\}, ..., \{v_n\})} \leq & \frac{1}{\sum_i f(\{u_i\})}, \label{eq:effub2}
\end{align}
which is true by Restrictions 1, 2 and 4. This shows how the values $f(\{u_1\}), ..., f(\{u_n\})$ directly determine the upper bound on efficiency of the algorithm for this scenario.

Since by definition $\gamma(f, X, G) \geq \gamma(G)$, one can find the lowest such upper bound on $\gamma(G)$ by setting $f(\{u_i\})$ so as to maximize $\sum_i f(\{u_i\})$ (from \eqref{eq:effub2}) subject to the constraint in \eqref{eq:effubconst}. This is precisely the linear program in \eqref{eq:fracindlp}, whose value is $\alpha^*(G)$. Therefore $1/{\alpha^*(G)}$ is an upper bound on $\gamma(G)$.

For the lower bound, let $x_i = x_i^{\rm sol}$ and $x = x^{\rm sol}$ for ease of notation. Then consider the following:
\begin{align}
f(x^{\rm opt}) \leq & f(x, x^{\rm opt}) = f(x) + \sum_i \Delta\left(x_i^{\rm opt} \Bigg| \bigcup_{j < i} x^{\rm opt}_j \cup x\right), \\
\leq & f(x) + \sum_i \Delta(x^{\rm opt}_i | x_{\mathcal{N}_i}), \label{eq:lbstart2}\\
\leq & f(x) + \sum_i \Delta(x_i | x_{\mathcal{N}_i}), \label{eq:lbstart}
\end{align}
where \eqref{eq:lbstart2} follows from submodularity and \eqref{eq:lbstart} follows from the decision-making rule in \eqref{eq:choice2}. Let $a \in K(G)$ and let $\beta_i^P = 1 - \sum_{c \in P:i \in c}y_c$ for $P \subseteq K(G), y_c \in \mathbb{R}$. Then it follows that
\begin{align}
f(x^{\rm opt}) \leq & f(x) + \sum_{i \in a} \Delta(x_i | x_{\mathcal{N}_i}) + \sum_{i \notin a} \Delta(x_i | x_{\mathcal{N}_i}), \label{eq:lbmid1}\\
\leq & f(x) + y_a\sum_{i \in a} \Delta(x_i | x_{\mathcal{N}_i \cap a}) \label{eq:lbmid2}\\
& + (1-y_a)\sum_{i \in a} \Delta(x_i | x_{\mathcal{N}_i}) + \sum_{i \notin a} \Delta(x_i | x_{\mathcal{N}_i}),  \nonumber \\
= & f(x) + y_af(x_a) + \sum_{i} \beta_i^{\{a\}} \Delta(x_i |x_{\mathcal{N}_i}) \label{eq:lbmid3}
\end{align}
where \eqref{eq:lbmid2} is true by submodularity and \eqref{eq:lbmid3} is true by \eqref{eq:margclique} with some algebraic manipulation. The procedure followed in \eqref{eq:lbmid1}--\eqref{eq:lbmid3} can be thought of as an algorithm to put the sum in \eqref{eq:lbstart} into a more convenient form (for our purposes) with respect to some clique $a$. One could run the same procedure on the sum in \eqref{eq:lbmid3} with respect to some clique $b$, to obtain:
\begin{align}
f(x^{\rm opt}) \leq & f(x) + \sum_{c \in \{a, b\}} y_cf(x_c) + \sum_{i} \beta_i^{\{a, b\}}\Delta(x_i |x_{\mathcal{N}_i}).
\end{align}
Adding more cliques to the set $\{a, b\}$ simply requires an update on the two places that $\{a, b\}$ appears in the equation. Therefore, if the set of cliques is the full set $K(G)$, then
\begin{equation}
\label{eq:lbmidmed}
f(x^{\rm opt}) \leq f(x) + \sum_{c \in K(G)} y_cf(x_c) + \sum_i \beta_i^{K(G)} \Delta(x_i | x_{\mathcal{N}_i}).
\end{equation}
Note that this holds for any set of $y_c$, but in order to remove the third term from the inequality, we impose a constraint that $\beta_i^{K(G)} \leq 0$. Alternatively stated, we require that
\begin{equation}
\label{eq:lbconst}
\sum_{c \in K(G):i \in c} y_c \geq 1\ \  \forall i.
\end{equation}
Under these conditions and by monotonicity, \eqref{eq:lbmidmed} becomes $f(x^{\rm opt}) \leq f(x) + f(x) \sum_{c \in K(G)} y_c$, which implies that
\begin{equation}
\label{eq:efflb}
\gamma(G) \geq \frac{1}{1 + \sum_{c \in K(G)} y_c}.
\end{equation}
To find the highest such lower bound on $\gamma(G)$, one needs to find the minimum value for the sum in \eqref{eq:efflb}, subject to the constraint in \eqref{eq:lbconst}. This is precisely the linear program defined in \eqref{eq:fraccliquecovlp}. Therefore, it follows that $\gamma(G) \geq 1/(1 + k^*(G)) = 1/(1 + \alpha^*(G))$. \qed

%% file: src/structures.tex
\section{Optimal Structures}
\label{sec:struct}

In this section, we describe how to build a graph $G$ that yields the highest efficiency $\gamma(G)$ subject to a constraint on the number of edges.

\subsection{Preliminaries}

\begin{figure}
	\centering
	\begin{subfigure}[b]{0.48\textwidth}
		\centering
		\includegraphics[scale=0.5]{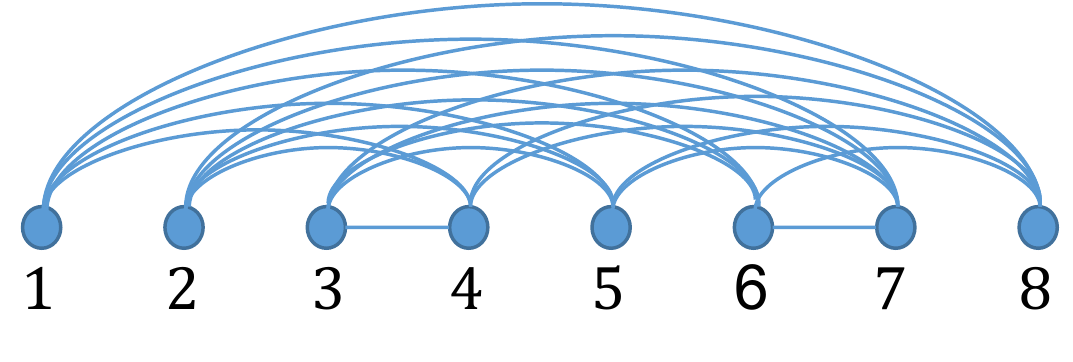}
		\caption{The Tur\'{a}n graph $T(8, 3)$, where $\omega(G)=3$. No other graph with 8 vertices can have more edges without also having a clique of size 4 or higher.\\}
	\end{subfigure}
	\begin{subfigure}[b]{0.48\textwidth}
		\centering
		\includegraphics[scale=0.5]{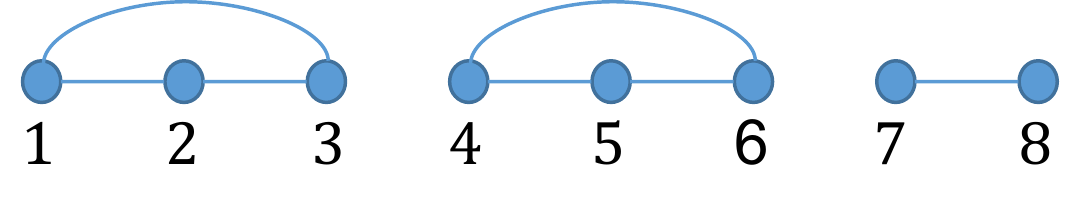}
		\caption{The complement Tur\'{a}n graph $\overline{T(8, 3)}$, where $\alpha(G)=3$. No other graph with 8 vertices can have fewer edges without also having an independent set of size 4 or higher.}
		\label{fig:turancomp}
	\end{subfigure}
	\caption{A Tur\'{a}n graph and its complement}
	\label{fig:turan}
\end{figure}

We denote $\mathcal{G}_{m, n} := \{G = (V, E) \in \mathcal{G} : |V| = n, |E| \leq m\}$ and $G^*_{m,n} \in \argmax_{G \in \mathcal{G}_{m,n}} \gamma(G)$, i.e., $G^*_{m,n}$ is a graph in $\mathcal G_{m,n}$ that maximizes efficiency. The complement $\bar{G} = (\bar{V}, \bar{E})$ of graph $G = (V, E)$ is such that $\bar{V} = V$ and $(i, j) \in \bar{E}$ if and only if $(i, j) \notin E$. It is straightforward to show that $\alpha(G) = \omega(\bar{G})$.

In graph theory a \emph{Tur\'{a}n graph} $T(n,r)$ is a graph with $n$ vertices created with the following algorithm:
\begin{enumerate}
	\item Partition the vertices into $r$ disjoint sets $C_1, ..., C_r$ such that $|C_i| - |C_j| \leq 1$ for all $i,j \in \{1, ..., r\}$.
	\item Create edges between all nodes not within the same set.
\end{enumerate}
A result known as Tur\'{a}n's theorem states that $T(n, r)$ is an $n$-node graph with the highest number of edges that has clique number $r$ or smaller~\cite{turan1941extremal}. Alternatively stated,
\begin{equation}
\label{eq:turanthm}
T(n, r) \in \argmax_{G=(V,E) : \omega(G) \leq r} |E|.
\end{equation}

\begin{figure*}
	\centering
	\includegraphics[scale=0.45]{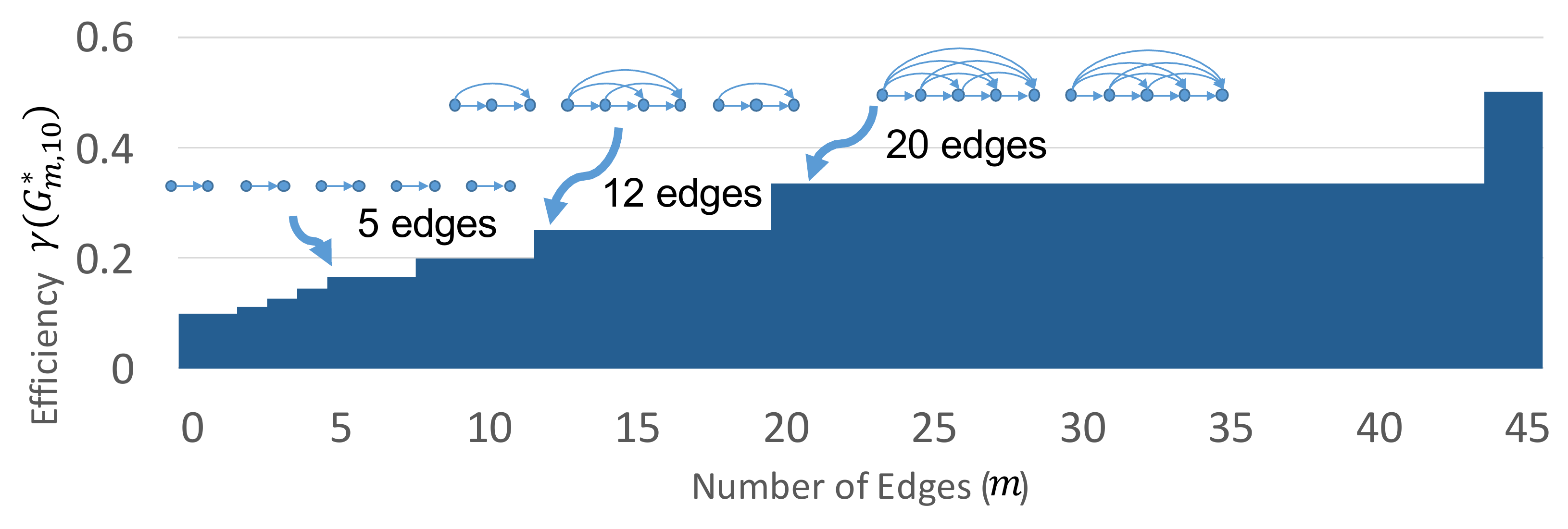}
	\caption{The efficiency of $G^*_{m,10}$ for all values of $m$, with example graphs for a few values of $m$. Notice the ``dead zones", where adding more edges does not lead to any higher efficiency guarantees.}
	\label{fig:struct_chart}
\end{figure*}

The complement of a Tur\'{a}n graph, denoted $\overline{T(n, r)}$, is created with the same procedure as a Tur\'{a}n graph, except that in Step 2, edges are created among all nodes \emph{within} the same set. An example of a Tur\'{a}n graph and its complement is found in Figure \ref{fig:turan}. Thus we can also state, similar to \eqref{eq:turanthm}, that
\begin{equation}
	\label{eq:min_ind}
	\overline{T(n, r)} \in \argmin_{G=(V,E) : \alpha(G) \leq r} |E|.
\end{equation}
In words $\overline{T(n, r)}$ is a graph with the fewest edges that has independence number $r$. It should also be clear that 
\begin{equation}
\label{eq:turanalphak}
	\alpha\left(\overline{T(n,r)}\right) = \alpha^*\left(\overline{T(n, r)}\right) = k\left(\overline{T(n, r)}\right)=r.
\end{equation}
Lastly, we define the graph
\begin{equation}
	\hat{T}(n, m) := \argmin_{\left\{\overline{T(n,r)}: |E| \leq m \right\}} r,
\end{equation}
which is the complement $n$-node Tur\'{a}n graph with the lowest independence number among all graphs with the number of edges less than or equal to $m$ \footnote{Searching over the space of complement Tur\'{a}n graphs can be done simply. Adapting part of Tur\'{a}n's theorem, we see that $r \geq \ceil{n^2/(2m+n)}$. Therefore, one can start by setting $r$ to this minimum value, and then determining whether $m \geq M(n, r)$, see Lemma \ref{lem:turanedges} below. If the statement is not true, $r$ can be incremented until it is.}.

\subsection{Result}

The main result of this section regarding efficient graph structures is stated below and later proved in Section \ref{subsec:thmstruct}.

\begin{theorem}
	\label{thm:struct}
	Consider two nonnegative integers $n$ and $m$ such that $m \leq \frac{1}{2}n(n-1)$. If $m \neq \frac{1}{2}n(n-1) - 1$, then $G^*_{m,n} = \hat{T}(n, m)$. If $m = \frac{1}{2}n(n-1) - 1$, then $G^*_{m,n}$ is the full clique on $n$ nodes, minus the edge $(n-1, n)$. 
\end{theorem}

An illustration of $\gamma(G^*_{m,n})$ as a function of the number of edges $m$ is given in Figure \ref{fig:struct_chart}. One item to note is that there may be extra edges not used in our design of $G^*_{m,n}$. For instance, in Figure \ref{fig:struct_chart}, the efficiency is the same when $12 \leq m \leq 19$. This implies that $G^*_{12, 10}$ and $G^*_{19, 10}$ can be the same graph, and for any value of $m$ in between. Hence, there are ``dead zones" seen in the graph in Figure \ref{fig:struct_chart}.

\subsection{The Sibling Property}
\label{subsec:sp}

Here we present a graph property, along with a corollary to Theorem \ref{thm:eff}. These results are key to the proof for Theorem \ref{thm:struct}.

\begin{defn}
	\label{def:sp}
	Let $G \in \mathcal{G}$. Then $G$ has the \emph{Sibling Property} if for some maximum independent set $J$, there exist $w \in V \setminus J$ and $i \in J$ such that $i \in \mathcal{N}_w$ (see Figure \ref{fig:ex}).
\end{defn}

\begin{lemma}
	\label{lem:spprop}
	If a graph $G$ lacks the Sibling Property, then
	\begin{enumerate}
		\item \label{itm:uniquemax} There is a unique maximum independent set $J$.
		\item \label{itm:lastnode} The set $J$ must include nodes $n$ and $n-1$.
		\item \label{itm:indsub} The induced subgraph $G'$ created by removing the set $J$ from $G$ must be such that $\alpha(G) > \alpha(G')$.
		\item \label{itm:conn} Every node outside the set $J$ must have outgoing edges to at least 2 nodes in $J$ \footnote{In the literature, such a $J$ is called a \emph{perfect independent set}. We present a proof here that suits the needs of this work, but it is also shown in \cite{volkmann2004perfect} that every unique maximum independent set is perfect.}.
	\end{enumerate}
	\begin{proof}
		We prove each Propoerty separately:
		
		\emph{Property \ref{itm:uniquemax}:} Suppose there are 2 maximum independent sets $J$ and $J'$. Let $i \in J' \setminus J$ and $j \in J \setminus J'$. By definition, all nodes in either set cannot have any outgoing edges. This implies that $(i, j), (j, i) \notin E$: in other words, $i$ and $j$ are independent from each other, and neither $J$ nor $J'$ are maximum, a contradiction. 
		
		\emph{Property \ref{itm:lastnode}:} First, suppose that $n$ is not included in $J$. Then there exists an edge $(i, n)$ for some $i \in J$. By defintion, this means $G$ has the Sibling Property, a contradiction. Now suppose that $n-1$ is not in $J$. Since $G$ does not have the Sibling Property, then by definition $(j, n-1) \notin E$ for all $j \in J \setminus n$. This means that another maximum independent set is $\{n-1\} \cup J \setminus n$, which is a contradiction to statement \ref{itm:uniquemax}. 
		
		\emph{Property \ref{itm:indsub}:} If this were not true, then $J$ would not be a unique maximum independent set. 
		
		\emph{Property \ref{itm:conn}:} Let $i \notin J$. By definition, $i$ cannot have any incoming edges from $J$ and if there are no edges between $i$ and $J$, then $i$ must be part of $J$, a contradiction. Therefore, we consider the case where $(i, j) \in E$ for some $j \in J$, but no outgoing edges from $i$ to $J \setminus j$ exist. This means that another maximum independent set is $i \cup J \setminus j$, and $J$ is not unique. By Property \ref{itm:uniquemax}, this is a contradiction.
	\end{proof}
\end{lemma}

\begin{corollary}
	\label{cor:effsp}
	For a graph $G \in \mathcal{G}$ with the Sibling Property,
	\begin{equation}
	\label{eq:effsp}
	\gamma(G) \leq \frac{1}{1 + \alpha(G)},
	\end{equation}
	with equality when $\alpha(G) = k(G)$
	\begin{proof}
		We provide an example which gives us the upper bound using a weighted set cover problem. Let $J$ be a maximum independent set of $G$ and let $w$ be defined as in Definition \ref{def:sp}. Then $\mathcal{T} = \{t_1, ..., t_n\}$, where $v_i = 1$ if $i \in J$ or $i = w$, and $v_i = 0$ otherwise. The action sets are
		\begin{equation}
			X_i = \left \{
			\begin{array}{ll}
				\{\{t_w\}, \{t_i\}\} & \mbox{if $i \in J$,}\\
				\{\{t_w\}\} & \mbox{if $i = w$} \\
				\{\{t_i\}\} & \mbox{otherwise.}
			\end{array} 
			\right.
		\end{equation}
		
		Each agent in $J$ is equally incentivized to choose either option, since none of them can access to the choice of the others. Therefore, the worst case in the greedy algorithm is for every agent in $J$ to choose $t_w$, implying $f(x^{\rm sol}) = v_w = 1$. Each agent makes the other choice in the optimal, so $f(x^{\rm opt}) = v_w + \sum_{i \in J} v_i = 1 + \alpha(G)$. Therefore $\gamma(f, X, G) = 1/(1 + \alpha(G))$ is an upper bound on $\gamma(G)$.
		
		In the case where $\alpha(G) = k(G)$, \eqref{eq:fracprop} shows that $\alpha^*(G) = \alpha(G)$, which implies by Theorem \ref{thm:eff} that $\gamma(G) \geq 1/(1 + \alpha(G))$.
	\end{proof}
\end{corollary}

\subsection{Proof for Theorem \ref{thm:struct}}
\label{subsec:thmstruct}

In this section we present the proof for Theorem \ref{thm:struct}, beginning with two lemmas. The first characterizes the number of edges in a complement Tur\'{a}n graph, and the second characterizes the fewest number of edges in a graph without the Sibling Property.

\begin{lemma}
	\label{lem:turanedges}
	Let $G = \overline{T(n, r)}$. Then the number of edges in $G$ is
	\begin{align}
	\label{eq:turanedges}
	M(n, r) :=& \frac{1}{2}(n \bmod r) \ceil*{\frac{n}{r}}\left(\ceil*{\frac{n}{r}} - 1\right) + \nonumber \\ & \frac{1}{2}(r - n \bmod{r}) \floor*{\frac{n}{r}} \left(\floor*{\frac{n}{r}} -1\right).
	\end{align}
	\begin{proof}
		This can be shown by construction. Recall that $\overline{T(n, r)}$ is a set of disconnected cliques, of as close to equal size as possible. Making purely equal-sized cliques would mean that each clique is of size $\floor{n/r}$, with $n \bmod r$ nodes left over. If each of these remaining nodes is added to a different clique, then $G$ consists of $n \bmod r$ cliques of size $\ceil{n/r}$ and the rest of size $\floor{n/r}$. Since a clique of size $p$ contains $\frac{1}{2}p(p-1)$ edges, we can see that the first line in \eqref{eq:turanedges} is the number of edges in all the larger cliques, and the second line is the number of edges in all the smaller cliques.
	\end{proof}
\end{lemma}

\begin{lemma}
	\label{lem:nospedges}
	Let $G \in \mathcal{G}$ have $n$ nodes, be without the Sibling Property, and such that $\alpha(G) = r$. Then the number of edges $m$ in $G$ satisfies
	\begin{equation}
	\label{eq:nospedges}
	m \geq M(n-r, r-1) + 2(n-r).
	\end{equation}
	Furthermore, for any values of $n$ and $r$, such a $G$ can be constructed so that \eqref{eq:nospedges} is at equality.
	\begin{proof}
		In this proof, we construct a $G$ such that \eqref{eq:nospedges} is at equality, then reason that no other graph with $n$ nodes, independence number $r$, and without the Sibling Property can have fewer edges. The proof also leverages the properties for a graph without the Sibling Property, found in Lemma \ref{lem:spprop}. Let $J$ be the unique maximum independent set (Property \ref{itm:uniquemax}) in $G$, and let $G'$ be the induced subgraph of $G$ created by removing the nodes in $J$. Then we know that $\alpha(G') < \alpha(G)$ (Property \ref{itm:indsub}). From \eqref{eq:min_ind} and Lemma \ref{lem:turanedges}, the minimum number of edges that such a $G'$ can have is $M(n-r, r-1)$. Finally, every node in $G'$ must have outgoing edges to at least two nodes in $J$, therefore, $G$ must have an additional $2(n-r)$ edges. Thus the minimum number of edges to construct $G$ is given in \eqref{eq:nospedges}.
	\end{proof}
\end{lemma}

We now commence with the proof for Theorem \ref{thm:struct}. The case $m=0$ trivially holds, so we assume that $m>0$. Recall that the graph $\hat{T}(n, m)$ is a set of disconnected cliques, which implies that any maximum independent set has one node from each clique, and that no maximum independent set is unique. Therefore, by Lemma \ref{lem:spprop}, Property \ref{itm:uniquemax}, $\hat{T}(n, m)$ has the Sibling Property. In light of \eqref{eq:turanalphak}, It follows from Corollary \ref{cor:effsp} that $\gamma(\hat{T}(n, m)) = 1/(1 + \alpha(\hat{T}(n, m))$. The statement in \eqref{eq:min_ind} also shows that no other graph with $\leq m$ edges can have a smaller independence number. Combining this with Corollary \ref{cor:effsp} implies that no other graph with the Sibling Property (and same number of nodes and edges) can have a higher efficiency.

It remains to confirm that any graph without the Sibling Property cannot have a higher efficiency than $\hat{T}(n,m)$, given $n$ nodes and $m$ edges -- with the exception when $m = \frac{1}{2}n(n-1)$. Let $G$ be a graph with $n$ nodes, $m$ edges, without the Sibling Property, and with independence number $r+1$. We assume that $G$ has the fewest number of edges (as dictated by Lemma \ref{lem:nospedges}), with the highest possible efficiency $\gamma(G) = 1/(r+1)$. By Corollary \ref{cor:effsp} and \eqref{eq:turanalphak}, this is the same efficiency as $\hat{T}(n, m)$, thus we seek to characterize when the number of edges in $G$ is greater than or equal to that of $\hat{T}(n, m)$. In other words, $\mathcal{G}^*_{m,n} = \hat{T}(n, m)$ only if
\begin{equation}
\label{eq:edgecomp2}
M(n, r) \leq m = M(n - r - 1, r) + 2(n - r - 1).
\end{equation}
In order to show when this condition holds, we divide the remainder of the proof into four cases, the union of which covers all possible values of $n$ and $r$. In the first case, when $r=1$, we prove \eqref{eq:edgecomp2} is false for all values of $n$ (which corresponds to the case in the theorem statement when $m = \frac{1}{2}n(n-1)-1$). In the other cases, we show that \eqref{eq:edgecomp2} is true, justifying that $G^*_{m,n} = \hat{T}(n,m)$.

\emph{Case 1: $r = 1$.} Here, $\hat{T}(n, m)$ is a clique, and has $\frac{1}{2}n(n-1)$ edges. The graph $G$ is such that
\begin{align}
m = & M(n-2, 1) + 2(n-2) = \frac{1}{2}n(n-1) - 1,
\end{align}
which is one less edge than $\hat{T}(n, m)$. Thus, for any value of $n$, there exists a $G$ where \eqref{eq:edgecomp2} is false. Such a $G$ is shown for $n=4$ in Figure \ref{fig:graph}, and a trivial extension to the proof in Appendix \ref{app:exupper} shows that $\gamma(G) = \gamma(\hat{T}(n, m)) = 1/2$ for any value of $n$. Since $G$ is created with the fewest number of edges, it follows that \eqref{eq:edgecomp2} is false \emph{only} when $m = \frac{1}{2}n(n-1) - 1$. By the construction in the proof of Lemma \ref{lem:nospedges}, such a $G$ is the full clique minus the edge $(n-1, n)$.

\emph{Case 2: $r = n-1 \geq 2$}. In this case, $\hat{T}(n,m)$ is the graph with no edges and efficiency $1/n$. Any graph with 1 or 0 edges must have this same efficiency, so \eqref{eq:edgecomp2} is true in this case.

In the remaining cases, we assume that $2 \leq r \leq n-2$, which also implies that $n \geq 4$. In both cases, we show that \eqref{eq:edgecomp2} holds.

\emph{Case 3: $n \bmod r \geq 1$.}  This condition implies the following:
\begin{itemize}
	\item $(n - r - 1) \bmod r = n \bmod r-1$
	\item $\ceil{n/r} = \floor{n/r} + 1$
	\item $\floor{(n-r-1)/r} = \floor{r/n} - 1$
	\item $\ceil{(n-r-1)/r} = \floor{r/n}$ 
\end{itemize}
Leveraging the above statements, $M(n, r)$ and $M(n-r-1,r)$ become:
\begin{align}
&M(n, r) = \frac{1}{2}\floor*{\frac{n}{r}} \bigg( 2(n \bmod r) + r \floor*{\frac{n}{r}} - r\bigg) \\
&M(n-r-1, r) = (n \bmod r )\left( \floor*{\frac{n}{r}} -1 \right) + \frac{r}{2}\floor*{\frac{n}{r}}^2 - 3\floor*{\frac{n}{r}}
\end{align}
Using these expressions to evaluate \eqref{eq:edgecomp2} yields
\begin{equation}
\label{eq:edgecomp4}
\floor*{\frac{n}{r}}(r+1) + n \bmod r \leq 2n - r - 1.
\end{equation}
We can now use the identity $n \bmod r = n - r\floor{n/r}$ to change the requirement in \eqref{eq:edgecomp4} to
\begin{align}
\floor*{\frac{n}{r}} \leq & n-r-1.
\end{align}
Since $\floor{n/r} \leq n/r$, a sufficient statement for \eqref{eq:edgecomp4} to hold can be found by replacing $\floor{n/r}$ with $n/r$, which can be simplified to
\begin{equation}
\label{eq:edgecomp5}
\frac{r^2 + 1}{r - 1} \leq n.
\end{equation}
The expression on the left side of the inequality is nondecreasing in $r$. Since $r \leq n-2$, if the inequality is true for $r = n-2$, then it is true for all relevant values of $n, r$. If we let $r = n-2$ in \eqref{eq:edgecomp5} and simplify, we conclude that \eqref{eq:edgecomp5} holds for all $n \geq 5$. By the premise that $n \geq 4$, the only values of $n,r$ that could make this false are $n=4, r=2$, however, this would imply that $n \bmod r = 0$, not allowable by the condition for this case. Thus, \eqref{eq:edgecomp5} is true and by extension so is \eqref{eq:edgecomp2}.

\emph{Case 4: $n \bmod r = 0$.} The condition implies the following:
\begin{itemize}
	\item $(n - r - 1) \bmod r = r-1$
	\item $\floor{n/r} = \ceil{n/r} = n/r$
	\item $\floor{(n-r-1)/r} = \floor{r/n} - 2$
	\item $\ceil{(n-r-1)/r} = \floor{r/n} - 1$
\end{itemize}
Leveraging the above statements, $M(n, r)$ and $M(n-r-1,r)$ become:
\begin{align}
&M(n, r) = \frac{n}{2} \left(\frac{n}{r} - 1\right), \\
&M(n-r-1, r) = \frac{n^2}{2r} - \frac{3}{2}n + r - \frac{n}{r} + 2.
\end{align}
Using these expressions to evaluate \eqref{eq:edgecomp2} yields
\begin{align}
\frac{r^2}{r-1}\leq & n. \label{eq:edgecomp6} 
\end{align}
It is straightforward to see that if \eqref{eq:edgecomp5} holds, then so does \eqref{eq:edgecomp6}. Case 3 shows that \eqref{eq:edgecomp5} is true unless $r=2, n=4$. However, \eqref{eq:edgecomp6} is true for these values, which implies \eqref{eq:edgecomp6} holds for all relevant values of $n, r$. Thus, in this case, \eqref{eq:edgecomp2} also holds.
 \qed

%% file: src/conclusion.tex
\section{Conclusion}

We have derived bounds on the worst-case efficiency of the distributed greedy algorithm for submodular maximization. These bounds can be used to design information sharing constraints that maximize the worst-case efficiency.

Future research can follow in several directions. For example, while the bounds presented in Theorem \ref{thm:eff} are applicable to any graph structure $G \in \mathcal{G}$, it is not immediately clear how to characterize graphs where $\gamma(G) = 1/\alpha^*(G)$ versus those where $\gamma(G) = 1/(\alpha^*(G) + 1)$, and if there exist graphs where $\gamma(G)$ is in between. Precisely defining $\gamma(G)$ in terms of its structural properties is another possibility.

Another idea for future research is to consider a game-theoretic approach instead of a greedy approach, similar to work done in \cite{vetta2002nash}. In this case, the order of the agents would not matter, and the goal would be to characterize the Nash equilibrium for certain graph structures. Finally, future work could include exploring the use of a different utility function rather than marginal contribution in order to make greedy decisions.

%% file: src/appendix.tex
\appendix

\subsection{Proof for Lower Bound in Example \ref{ex:upper}}
\label{app:exupper}

We begin with the following inequality:
\begin{align}
f(x^{\rm opt}) \leq & f(x^{\rm opt}, x_{1:2}) \label{eq:tightupper1}\\
= & f(x_{1:2}) + \Delta(x^{\rm opt}_1|x_{1:2}) \nonumber + \Delta(x^{\rm opt}_2 | x^{\rm opt}_1, x_{1:2}) \nonumber \\
& + \Delta(x^{\rm opt}_3 | x^{\rm opt}_{1:2}, x_{1:2}) + \Delta(x^{\rm opt}_4 | x^{\rm opt}_{1:3}, x_{1:2}) \label{eq:tightupper2}\\
\leq & f(x_{1:2}) + f(x_1^{\rm opt}) + \Delta(x_2^{\rm opt}|x_1) + \Delta(x_3^{\rm opt}|x_{1:2}) \nonumber \\
& + \Delta(x_4^{\rm opt} | x_{1:2}) \label{eq:tightupper3}\\
\leq & f(x_{1:2}) + f(x_1) + \Delta(x_2|x_1) + \Delta(x_3|x_{1:2}) \nonumber \\
&+ \Delta(x_4 | x_{1:2}) \label{eq:tightupper4}\\
=& f(x_{1:3}) + f(x_{1:2}, x_4) \leq  2f(x_{1:4}) \label{eq:tightupper5},
\end{align}
where \eqref{eq:tightupper1} is true by submodularity, \eqref{eq:tightupper2} is true by definition of $\Delta$, \eqref{eq:tightupper3} is true by submodularity, \eqref{eq:tightupper4} is true since agents choose according to \eqref{eq:choice2}, \eqref{eq:tightupper5} is true by definition of $\Delta$ and submodularity. Therefore we see $\gamma(G) = 1/2$. \qed

%% file: src/authors.tex
\begin{IEEEbiography}[{\includegraphics[width=1in,height=1.25in,clip,keepaspectratio]{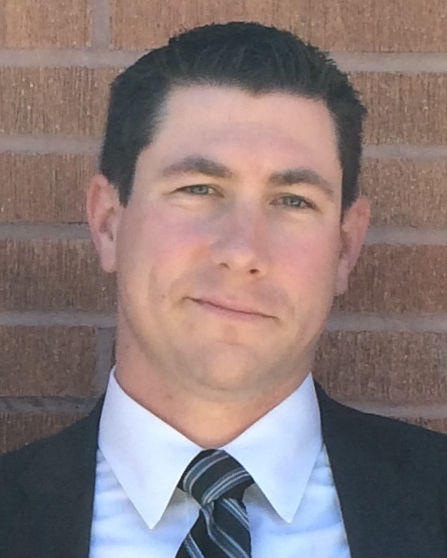}}]
	{David Grimsman} is a Ph.D. student in the Electrical and Computer Engineering Department at UC Santa Barbara. He completed BS in Electrical and Computer Engineering at Brigham Young University in 2006 as a Heritage Scholar, and with a focus on signals and systems. After working for BrainStorm, Inc. for several years as a trainer and IT manager, he returned to Brigham Young University and earned an MS in Computer Science in 2016. His research interests include mulit-agent systems, game theory, distributed optimization, network science, linear systems theory, and security of cyberphysical systems.
\end{IEEEbiography}
\begin{IEEEbiography}[{\includegraphics[width=1in,height=1.25in,clip,keepaspectratio]{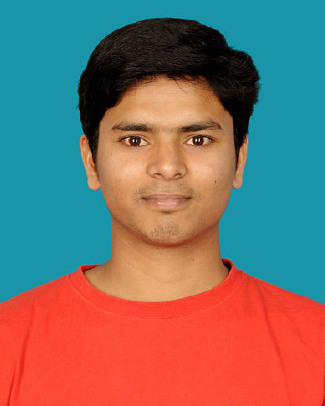}}]%
	{M. Shabbir Ali} is a postdoctoral researcher at Orange Labs, Paris. He did his Ph.D. in Computer Science and Networks from Telecom ParisTech, Paris in 2017. He received his M.Sc. degree in Electrical Communication Engineering from Indian Institute of Science, Bangalore in 2014. He did his Bachelor of Engineering degree in Electrical and Electronics Eng.\ from the MuffakhamJah College of Eng.\ and Technology, Hyderabad, India in 2010.  His research interests include game theory, machine learning, distributed optimization, heterogeneous cellular networks, cognitive radio networks, resource allocation, load balancing, and interference modeling and management.
\end{IEEEbiography}
\begin{IEEEbiography}[{\includegraphics[width=1in,height=1.25in,clip,keepaspectratio]{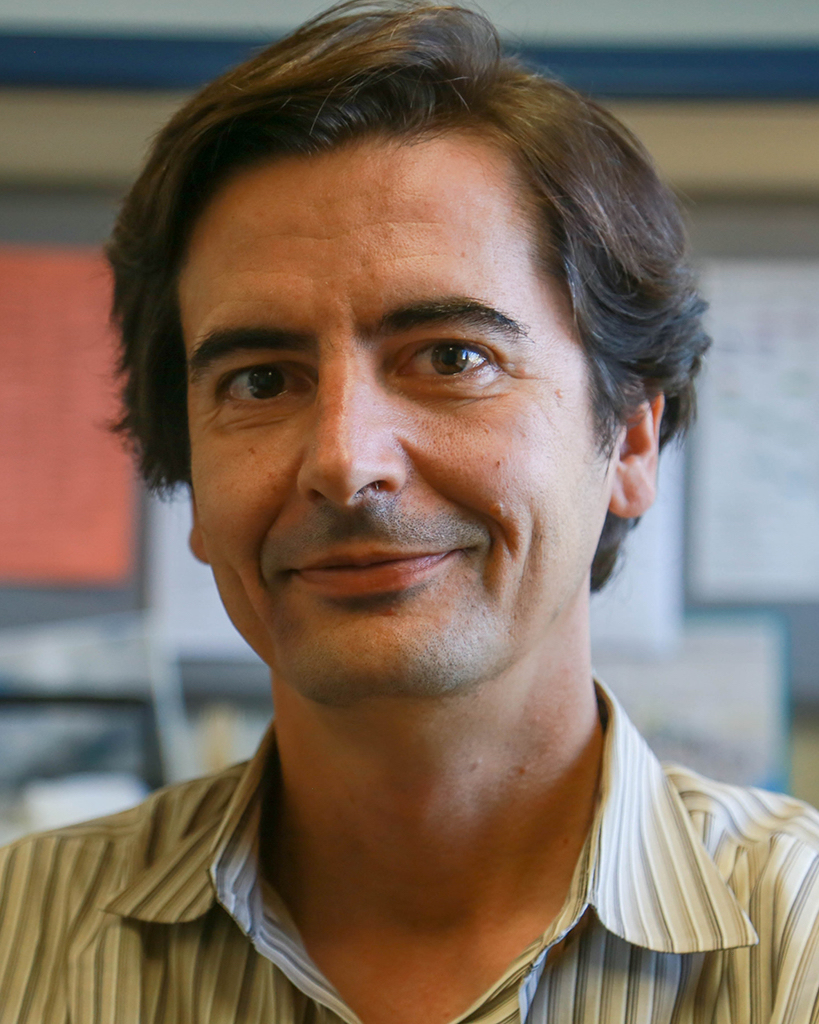}}]%
{Jo\~{a}o P. Hespanha} was born in Coimbra, Portugal, in 1968. He received the Licenciatura in electrical and computer engineering from the Instituto Superior Técnico, Lisbon, Portugal in 1991 and the Ph.D. degree in electrical engineering and applied science from Yale University, New Haven, Connecticut in 1998. From 1999 to 2001, he was Assistant Professor at the University of Southern California, Los Angeles. He moved to the University of California, Santa Barbara in 2002, where he currently holds a Professor position with the Department of Electrical and Computer Engineering.

Dr. Hespanha is the recipient of the Yale University’s Henry Prentiss Becton Graduate Prize for exceptional achievement in research in Engineering and Applied Science, a National Science Foundation CAREER Award, the 2005 best paper award at the 2nd Int. Conf. on Intelligent Sensing and Information Processing, the 2005 Automatica Theory/Methodology best paper prize, the 2006 George S. Axelby Outstanding Paper Award, and the 2009 Ruberti Young Researcher Prize. Dr. Hespanha is a Fellow of the International Federation of Automatic Control (IFAC) and of the IEEE. He was an IEEE distinguished lecturer from 2007 to 2013.

His current research interests include hybrid and switched systems; multi-agent control systems; game theory; optimization; distributed control over communication networks (also known as networked control systems); the use of vision in feedback control; stochastic modeling in biology; and network security.
\end{IEEEbiography}
\begin{IEEEbiography}[{\includegraphics[width=1in,height=1.25in,clip,keepaspectratio]{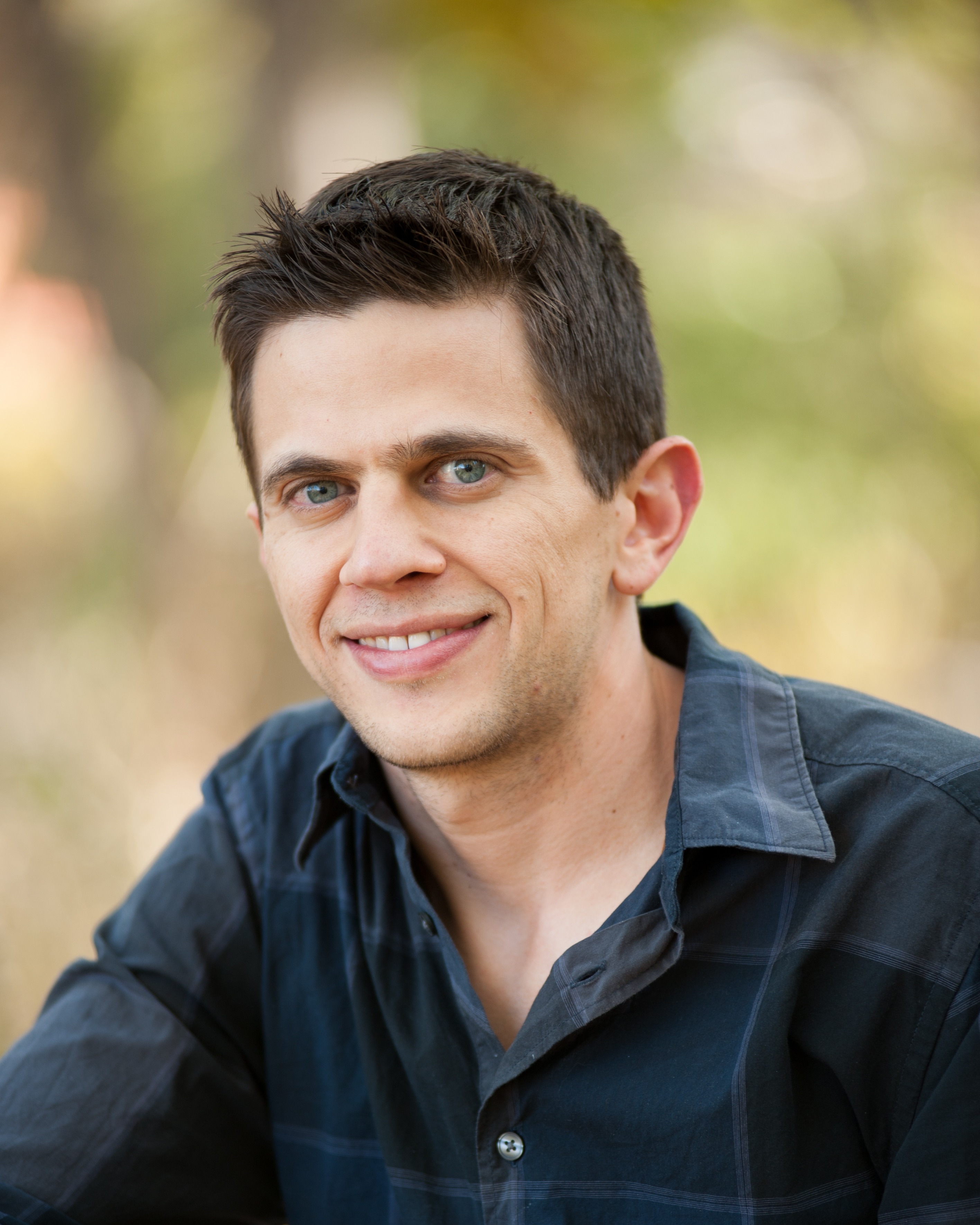}}]
{Jason R. Marden} is an assistant professor in the Department of Electrical and Computer Engineering at the University of California, Santa Barbara. He received the B.S. degree in 2001 and the Ph.D. degree in 2007 (under the supervision of Jeff S. Shamma), both in mechanical engineering from the University of California, Los Angeles, where he was awarded the Outstanding Graduating Ph.D. Student in Mechanical Engineering. After graduating, he was a junior fellow in the Social and Information Sciences Laboratory at the California Institute of  Technology until 2010 and then an assistant professor at the University of Colorado until 2015. He is a recipient of an ONR Young Investigator Award (2015), an NSF Career Award (2014), the AFOSR Young Investigator Award (2012), the SIAM CST Best Sicon Paper Award (2015), and the American Automatic Control Council Donald P. Eckman Award (2012). His research interests focus on game-theoretic methods for the control of distributed multiagent systems.
\end{IEEEbiography}

%% file: main.bbl
\begin{thebibliography}{10}

\bibitem{arslan2007autonomous}
G{\"u}rdal Arslan, Jason~R Marden, and Jeff~S Shamma.
\newblock Autonomous vehicle-target assignment: A game-theoretical formulation.
\newblock {\em Journal of Dynamic Systems, Measurement, and Control},
  129(5):584--596, 2007.

\bibitem{arumugam2007fractional}
Subramanian Arumugam and K.~Reji Kumar.
\newblock Fractional independence and fractional domination chain in graphs.
\newblock {\em AKCE International Journal of Graphs and Combinatorics},
  4(2):161--169, 2007.

\bibitem{barinova2012detection}
Olga Barinova, Victor Lempitsky, and Pushmeet Kholi.
\newblock On detection of multiple object instances using {H}ough transforms.
\newblock {\em IEEE TPAMI}, 34(9):1773--1784, 2012.

\bibitem{buchbinder2015tight}
Niv Buchbinder, Moran Feldman, Joseph Seffi, and Roy Schwartz.
\newblock A tight linear time (1/2)-approximation for unconstrained submodular
  maximization.
\newblock {\em SIAM Journal on Computing}, 44(5):1384--1402, 2015.

\bibitem{calinescu2007maximizing}
Gruia Calinescu, Chandra Chekuri, Martin P{\'a}l, and Jan Vondr{\'a}k.
\newblock Maximizing a submodular set function subject to a matroid constraint.
\newblock In {\em International Conference on Integer Programming and
  Combinatorial Optimization}, pages 182--196. Springer, 2007.

\bibitem{feige1998threshold}
Uriel Feige.
\newblock A threshold of ln $n$ for approximating set cover.
\newblock {\em Journal of the ACM (JACM)}, 45(4):634--652, 1998.

\bibitem{filmus2012power}
Yuval Filmus and Justin Ward.
\newblock The power of local search: Maximum coverage over a matroid.
\newblock In {\em Symposium on Theoretical Aspects of Computer Science
  (STACS)}, volume~14, pages 601--612. LIPIcs, 2012.

\bibitem{fisher1978analysis}
Marshall~L Fisher, George~L Nemhauser, and Laurence~A Wolsey.
\newblock An analysis of approximations for maximizing submodular set functions
  {II}.
\newblock In {\em Polyhedral Combinatorics}, pages 73--87. Springer, 1978.

\bibitem{gairing2009covering}
Martin Gairing.
\newblock Covering games: Approximation through non-cooperation.
\newblock In {\em International Workshop on Internet and Network Economics},
  pages 184--195. Springer, 2009.

\bibitem{gharesifard2016distributed}
B.~Gharesifard and S.~L. Smith.
\newblock Distributed submodular maximization with limited information.
\newblock {\em IEEE TCNS}, pages 1--1, 2017.

\bibitem{godsil2013algebraic}
Chris Godsil and Gordon~F. Royle.
\newblock {\em Algebraic graph theory}, volume 207.
\newblock Springer Science \& Business Media, 2013.

\bibitem{grimsmanimpact}
David Grimsman, Mohd~Shabbir Ali, Joao~P Hespanha, and Jason~R Marden.
\newblock Impact of information in greedy submodular maximization.
\newblock In {\em IEEE CDC 2017}, pages 2900--2905. IEEE, 2017.

\bibitem{grotschel1981ellipsoid}
Martin Gr{\"o}tschel, L{\'a}szl{\'o} Lov{\'a}sz, and Alexander Schrijver.
\newblock The ellipsoid method and its consequences in combinatorial
  optimization.
\newblock {\em Combinatorica}, 1(2):169--197, 1981.

\bibitem{iwata2001combinatorial}
Satoru Iwata, Lisa Fleischer, and Satoru Fujishige.
\newblock A combinatorial strongly polynomial algorithm for minimizing
  submodular functions.
\newblock {\em Journal of the ACM (JACM)}, 48(4):761--777, 2001.

\bibitem{kempe2003maximizing}
David Kempe, Jon Kleinberg, and {\'E}va Tardos.
\newblock Maximizing the spread of influence through a social network.
\newblock In {\em SIGKDD Conference on Knowledge Discovery and Data Mining},
  pages 137--146. ACM, 2003.

\bibitem{kohli2009p3}
Pushmeet Kohli, M.~Pawan Kumar, and Philip~H.S. Torr.
\newblock P$^3$ \& beyond: Move making algorithms for solving higher order
  functions.
\newblock {\em IEEE TPAMI}, 31(9):1645--1656, 2009.

\bibitem{krause2007near}
Andreas Krause and Carlos Guestrin.
\newblock Near-optimal observation selection using submodular functions.
\newblock In {\em Association for the Advancement of Artificial Intelligence},
  volume~7, pages 1650--1654, 2007.

\bibitem{krause2009simultaneous}
Andreas Krause, Ram Rajagopal, Anupam Gupta, and Carlos Guestrin.
\newblock Simultaneous placement and scheduling of sensors.
\newblock In {\em Proceedings of the 2009 International Conference on
  Information Processing in Sensor Networks}, pages 181--192. IEEE Computer
  Society, 2009.

\bibitem{lin2011class}
Hui Lin and Jeff Bilmes.
\newblock A class of submodular functions for document summarization.
\newblock In {\em ACL: Human Language Technologies}, volume~1, pages 510--520.
  Association for Computational Linguistics, 2011.

\bibitem{lovasz1983submodular}
L{\'a}szl{\'o} Lov{\'a}sz.
\newblock Submodular functions and convexity.
\newblock In {\em Mathematical Programming: The State of the Art}, pages
  235--257. Springer, 1983.

\bibitem{marden2016}
Jason~R. Marden.
\newblock The role of information in distributed resource allocation.
\newblock {\em IEEE TCNS}, 4(3):654--664, Sept 2017.

\bibitem{minoux1978accelerated}
Michel Minoux.
\newblock Accelerated greedy algorithms for maximizing submodular set
  functions.
\newblock In {\em Optimization Techniques}, pages 234--243. Springer, 1978.

\bibitem{mirzasoleiman2013distributed}
Baharan Mirzasoleiman, Amin Karbasi, Rik Sarkar, and Andreas Krause.
\newblock Distributed submodular maximization: Identifying representative
  elements in massive data.
\newblock In {\em Advances in Neural Information Processing Systems}, pages
  2049--2057, 2013.

\bibitem{nemhauser1978analysis}
George~L Nemhauser, Laurence~A Wolsey, and Marshall~L Fisher.
\newblock An analysis of approximations for maximizing submodular set functions
  {I}.
\newblock {\em Mathematical Programming}, 14(1):265--294, 1978.

\bibitem{qu2016distrib}
Guannan Qu, Dave Brown, and Na~Li.
\newblock Distributed greedy algorithm for mulit-agent task assignment problem
  with submodular utility functions.
\newblock Preprint submitted to Automatica, December 2016.

\bibitem{schrijver2000combinatorial}
Alexander Schrijver.
\newblock A combinatorial algorithm minimizing submodular functions in strongly
  polynomial time.
\newblock {\em Journal of Combinatorial Theory, Series B}, 80(2):346--355,
  2000.

\bibitem{singh2007efficient}
Amarjeet Singh, Andreas Krause, Carlos Guestrin, William~J Kaiser, and Maxim~A
  Batalin.
\newblock Efficient planning of informative paths for multiple robots.
\newblock In {\em International Joint Conferences on Artificial Intelligence},
  volume~7, pages 2204--2211, 2007.

\bibitem{sviridenko2004note}
Maxim Sviridenko.
\newblock A note on maximizing a submodular set function subject to a knapsack
  constraint.
\newblock {\em Operations Research Letters}, 32(1):41--43, 2004.

\bibitem{turan1941extremal}
Paul Tur{\'a}n.
\newblock On an extremal problem in graph theory.
\newblock {\em Mat. Fiz. Lapok}, 48(436-452):137, 1941.

\bibitem{vetta2002nash}
Adrian Vetta.
\newblock Nash equilibria in competitive societies, with applications to
  facility location, traffic routing and auctions.
\newblock In {\em Symposium on Foundations of Computer Science}, pages
  416--425. IEEE, 2002.

\bibitem{volkmann2004perfect}
Lutz Volkmann.
\newblock On perfect and unique maximum independent sets in graphs.
\newblock {\em Mathematica Bohemica}, 129(3):273--282, 2004.

\bibitem{vondrak2008optimal}
Jan Vondr{\'a}k.
\newblock Optimal approximation for the submodular welfare problem in the value
  oracle model.
\newblock In {\em ACM Symposium on Theory of Computing}, pages 67--74. ACM,
  2008.

\end{thebibliography}
